\documentclass[aps,prd,twocolumn,superscriptaddress]{revtex4-2}
\usepackage{graphicx}
\usepackage{amssymb}
\usepackage{amsmath}
\usepackage{booktabs}
\usepackage{multirow}
\usepackage{bm}
\usepackage{xcolor}
\usepackage{soul}
\usepackage{CJKutf8}
\definecolor{ceruleanblue}{rgb}{0.16, 0.32, 0.75}
\usepackage[colorlinks=true,linkcolor=ceruleanblue,citecolor=ceruleanblue,urlcolor=ceruleanblue]{hyperref}
\usepackage{tikz}
\usetikzlibrary{arrows.meta}
\usetikzlibrary{positioning}
\usetikzlibrary{calc}
\usetikzlibrary{decorations}
\usetikzlibrary{decorations.markings}
\tikzset{global scale/.style={scale=#1, every node/.append style={scale=#1}}}
\tikzset{
myarrowc/.style={ultra thick, postaction={decorate, decoration={markings, mark=at position 0.6 with {\arrow{latex}}}}},
myarrowl/.style={thick, postaction={decorate, decoration={markings, mark=at position 0.6 with {\arrow{latex}}}}},
}

\allowdisplaybreaks

\begin{document}
\title{Pion mass dependence in $D\pi$ scattering and the $D_0^*(2300)$ resonance from lattice QCD}
\author{Haobo Yan (\begin{CJK*}{UTF8}{gbsn}燕浩波\end{CJK*})}
\email{haobo@stu.pku.edu.cn}
\affiliation{School of Physics, Peking University, Beijing 100871, China}
\author{Chuan Liu (\begin{CJK*}{UTF8}{gbsn}刘川\end{CJK*})}
\email{liuchuan@pku.edu.cn}
\affiliation{School of Physics, Peking University, Beijing 100871, China}
\affiliation{Center for High Energy Physics, Peking University, Beijing 100871, China}
\affiliation{Collaborative Innovation Center of Quantum Matter, Beijing 100871, China}
\author{\mbox{Liuming Liu (\begin{CJK*}{UTF8}{gbsn}刘柳明\end{CJK*})}}
\email{liuming@impcas.ac.cn}
\affiliation{Institute of Modern Physics, Chinese Academy of Sciences, Lanzhou 730000, China}
\affiliation{University of Chinese Academy of Sciences, Beijing 100049, China}
\author{Yu Meng (\begin{CJK*}{UTF8}{gbsn}孟雨\end{CJK*})}
\email{yu\_meng@zzu.edu.cn}
\affiliation{School of Physics, Zhengzhou University, Zhengzhou 450001, China}
\author{Hanyang Xing (\begin{CJK*}{UTF8}{gbsn}邢瀚洋\end{CJK*})}
\affiliation{Institute of Modern Physics, Chinese Academy of Sciences, Lanzhou 730000, China}
\affiliation{University of Chinese Academy of Sciences, Beijing 100049, China}
\date{\today}

\begin{abstract}
Lattice QCD results for isospin $I=\frac{1}{2}$ $D\pi$ scattering are presented. Utilizing a series of $N_{\text{f}}=2+1$ Wilson-Clover ensembles with pion masses of $m_\pi \approx 133, 208, 305$ and $317$ MeV, various two-particle operators are constructed and the corresponding finite-volume spectra are determined. The $S$ and $P$-wave scattering phase shifts are extracted using the L\"{u}scher approach. A clear trend for the motion of the $D_0^*(2300)$ pole is identified. With the physical pion mass configurations also included, this calculation constitutes the first lattice calculation in which the pion mass dependence of the $D_0^*(2300)$ pole is investigated and the scattering lengths are extrapolated/interpolated to the physical pion mass in $D\pi$ scattering.
\end{abstract}

\maketitle

\section{Introduction}
\label{sec:intro}
Over two decades, both experimental and theoretical efforts have been devoted to the study of scalar resonance $D_0^*(2300)$, which has been discovered and verified by many experiments~\cite{Abe2004, Link2004, Aubert2009, Aaij2015, Aaij2015a}. Still, various puzzles remain. 
The scalar resonance $D_0^* (2300)$ lies so close to its strange partner $D_{s0}^* (2317)$, which is attributed within the quark model to the strong coupling to the $D\pi$ threshold~\cite{Chen2017}. There are also suggestions of a two-pole structure in this particular channel from unitarized chiral perturbation theory~\cite{Albaladejo2017, Asokan2023}. 
However, this assertion needs to be tested at different quark masses, which is impossible in real-world experiments. It is, therefore, crucial to study $D\pi$ scattering systematically within a non-perturbative framework such as lattice QCD, in which pion mass dependence could be analyzed. This is what we would like to address in this Letter. 

The study of the $D\pi$ two-meson system also provides a pivotal input to the more complicated three-body $DD\pi$ system. An example that has drawn much attention recently is the doubly-charmed tetraquark candidate, $T_{cc}(3875) \to DD^* \to DD\pi$ discovered by the LHCb collaboration~\cite{Aaij2022a, Aaij2022}. There have been several lattice calculations on $DD^*$ scattering recently, claiming the state that is related to $T_{cc}$ could be identified~\cite{Padmanath2022, Chen2022, Lyu2023, Collins2024}. However, the actual lowest $T_{cc}$ decay channel is a three-body channel and the large pion mass used in these lattice studies might obscure such complexity by lifting the $DD\pi$ channel beyond the $DD^*$ threshold. The more complete approach thus involves the three-body scattering problem on the lattice which relies on the $D\pi$ two-body scattering amplitudes as an input~\cite{Hansen2024, Dawid2024}.

To search for hadronic resonances on the lattice, the scattering amplitudes of the decay products are obtained by applying the L\"{u}scher's formulation~\cite{Luescher1991} to the finite-volume energy levels. The scattering amplitudes are then analytically continued to the complex energy plane. The poles found in these scattering amplitudes correspond to various hadrons found in the experiments. For a review, see~\cite{Mai2023}. There are, however, indirect extraction of the $I=\frac{1}{2}$ $D\pi$ scattering parameters. In Ref.~\cite{Flynn2007}, the scattering length is extracted from the lattice-determined form factor by a combined fit of the scattering length and the Omn\`{e}s subtraction parameters. In Ref.~\cite{Liu2013} the scattering length of $I=\frac{1}{2}$ $D \pi$ at physical $m_{\pi}$ was obtained within chiral perturbation theory with low-energy constants determined from other channels. There are also studies from phenomenological models, \textit{e.g.}, Ref.~\cite{Meng2023, Du2018, Guo2009, Guo2019, Huang2022, Guo2018, Guo2022, Korpa2023, Lutz2022}.

Direct lattice QCD calculations on $I=\frac{1}{2}$ $D\pi$ scattering were conducted on an $N_{\text{f}}=2$, $m_{\pi} \approx 266$ MeV lattice~\cite{Mohler2013} in which a resonance pole was found. Afterwards, the Hadron Spectrum Collaboration performed a $D\pi, D\eta, D_s\bar{K}$ coupled-channel analysis on an $N_{\text{f}}=2+1$ anisotropic lattice ensemble at $m_{\pi} \approx 391$ MeV and found $D_0^*$ to be a bound state~\cite{Moir2016}. They also performed a single-channel analysis at $m_{\pi} \approx 239$ MeV and found a resonance~\cite{Gayer2021}. A calculation at the $SU(3)$ point is also performed~\cite{Yeo2024}. However, focusing on the physical strange sector, the pole positions determined by the lattice studies are all lower than the PDG average value~\cite{Workman2022}. As $m_{\pi}$ decreases towards its physical value, although the scattering lengths appear to move towards the values determined from the experimental results, the pole position derived from these lattice results does not show a monotonic movement. Therefore, it is crucial to examine the movement of the $D_0^*$ pole at different $m_{\pi}$ values, particularly its physical value. Additionally, the experimental measurements are presented as the parameters of the relativistic Breit-Wigner (RBW), which could be rather different from the pole positions for broad resonances such as $D_0^*(2300)$. A more consistent comparison is to translate not only the lattice results but also those from the experiments to the pole positions.

In this Letter, we report lattice results on $D\pi$ scattering using a set of $N_{\text{f}}=2+1$ Wilson-clover lattice ensembles at $m_{\pi} \approx 133, 208, 305$ and $317$ MeV to examine the movement of the $D_0^*$ pole in the complex plane. The scattering lengths are interpolated/extrapolated towards the physical pion mass point. The pole positions and the scattering lengths are then compared to the experimental measurements. The behavior of the $D_0^*$ pole exhibits an interesting trend. At heavy $m_{\pi}$, the $D_0^*$ initially appears as a bound state, becoming lighter as $m_{\pi}$ decreases and transforming into a virtual state. With further reduction in $m_{\pi}$, the $D_0^*$ transitions into a resonance, showing a complicated movement.

\section{Finite volume spectra}
\label{sec:spectra}
The gauge configurations used in this work were generated by the CLQCD Collaboration with $N_{\text{f}} = 2+1$ flavors of dynamical quarks using the tadpole-improved tree-level Symanzik gauge action and the tadpole-improved tree-level Clover fermions~\cite{Hu2024}. The action of the valence charm quark is the same as the light and strange quark action used in the configurations. Furthermore, in order to investigate the discretization error on the coarsest lattice C48P14 with $m_{\pi} \approx 133$ MeV, we repeat the analysis using the Fermilab action~\cite{ElKhadra1997} for the charm quark, which controls discretization errors of $\mathcal{O}(am_c)^n$. The tuning of the parameters in the Fermilab action follows the method applied in Ref.~\cite{Liu:2009jc}. There are plenty of studies based on these ensembles, see, for \textit{e.g.}, ~\cite{Zhang2022, Xing2022, Liu2023, Liu2024, Liu2024a, Yan2024a, Meng2024, Meng2024a, Meng2024b, Han2024, Yan2024b}. The simulations were performed on six ensembles with four different values of pion mass  $m_\pi \approx 133, 208, 305$ and $317$ MeV. The details of the ensembles are listed in Tab.~\ref{tab:ens}. Among these ensembles, F32P21/F48P21 and F32P30/F48P30 are two couples that share the same pion mass and lattice spacing but different volumes to obtain more kinematic points in the finite-volume spectra, rendering a more stable and precise determination of the scattering parameters.

\begin{table}[htbp]
\caption{Gauge configurations used in this work. The errors are purely statistical.}
\begin{tabular}{ccccc}
\toprule
configuration & volume & $a$/fm & $m_{\pi}$/MeV & $N_{\text{confs}}$ \\
\midrule
C48P14 & $48^3 \times 96$ & $0.10530(18)$ & $133.1(1.6)$ & $259$ \\
F32P21 & $32^3 \times 64$ & $0.07746(18)$ & $206.8(2.1)$ & $459$ \\
F48P21 & $48^3 \times 96$ & $0.07746(18)$ & $208.12(70)$ & $222$ \\
F32P30 & $32^3 \times 96$ & $0.07746(18)$ & $305.81(71)$ & $567$ \\
F48P30 & $48^3 \times 96$ & $0.07746(18)$ & $304.98(50)$ & $201$ \\
H48P32 & $48^3 \times 144$ & $0.05187(26)$ & $317.00(68)$ & $274$ \\
\bottomrule
\end{tabular}
\label{tab:ens}
\end{table}

To obtain the full finite-volume spectra hence the scattering information near the threshold, many operators in irreducible representations (irreps) of the cubic space group are needed. We consider reference frames with a center of mass (CM) three-momenta $\frac{L}{2\pi} |\vec{P}| \leq 2$.

The finite-volume spectra can be obtained from the matrix of the correlation functions of the interpolating operators $O_i$, $C_{i j}(t)= \sum_{t^{\prime}} \langle O_i(t+t') O_j^{\dagger}(t^{\prime}) \rangle_T$, where $t^{\prime}$ is summed over all time slices to enhance the signal. The distillation quark smearing method~\cite{Peardon2009} is used to compute the quark propagators. The details about this method and the explicit forms of operators used in this work can be found in Ref.~\cite{supp}, where the strategies proposed in Ref.~\cite{Dudek2010, Thomas2012, Prelovsek2017} are applied. The construction has been performed with the Mathematica tool \verb|OpTion|~\cite{Yan2024}.

\begin{figure*}[htbp]
\includegraphics[height=0.25\textheight]{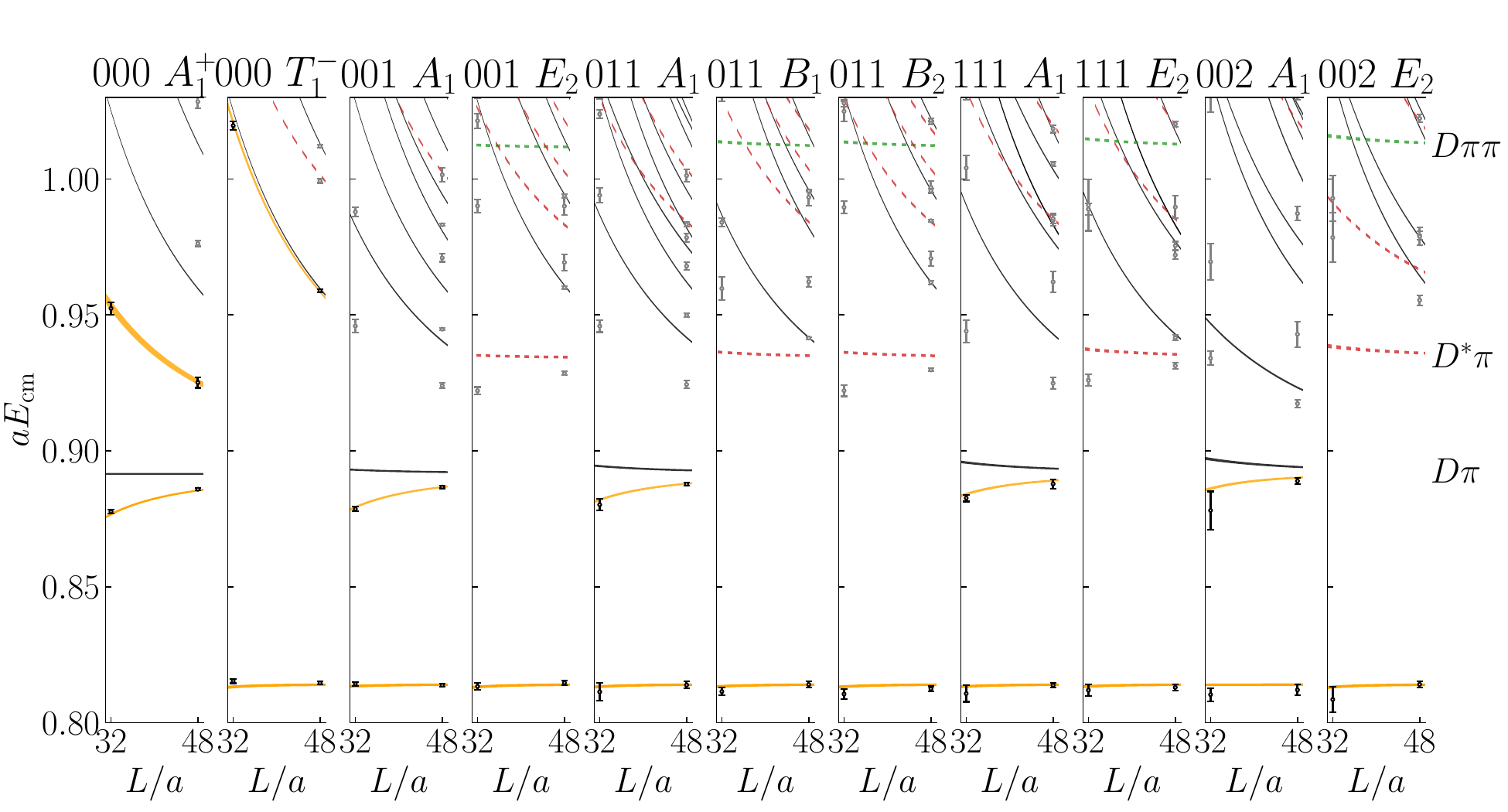}
\hfill
\includegraphics[height=0.25\textheight]{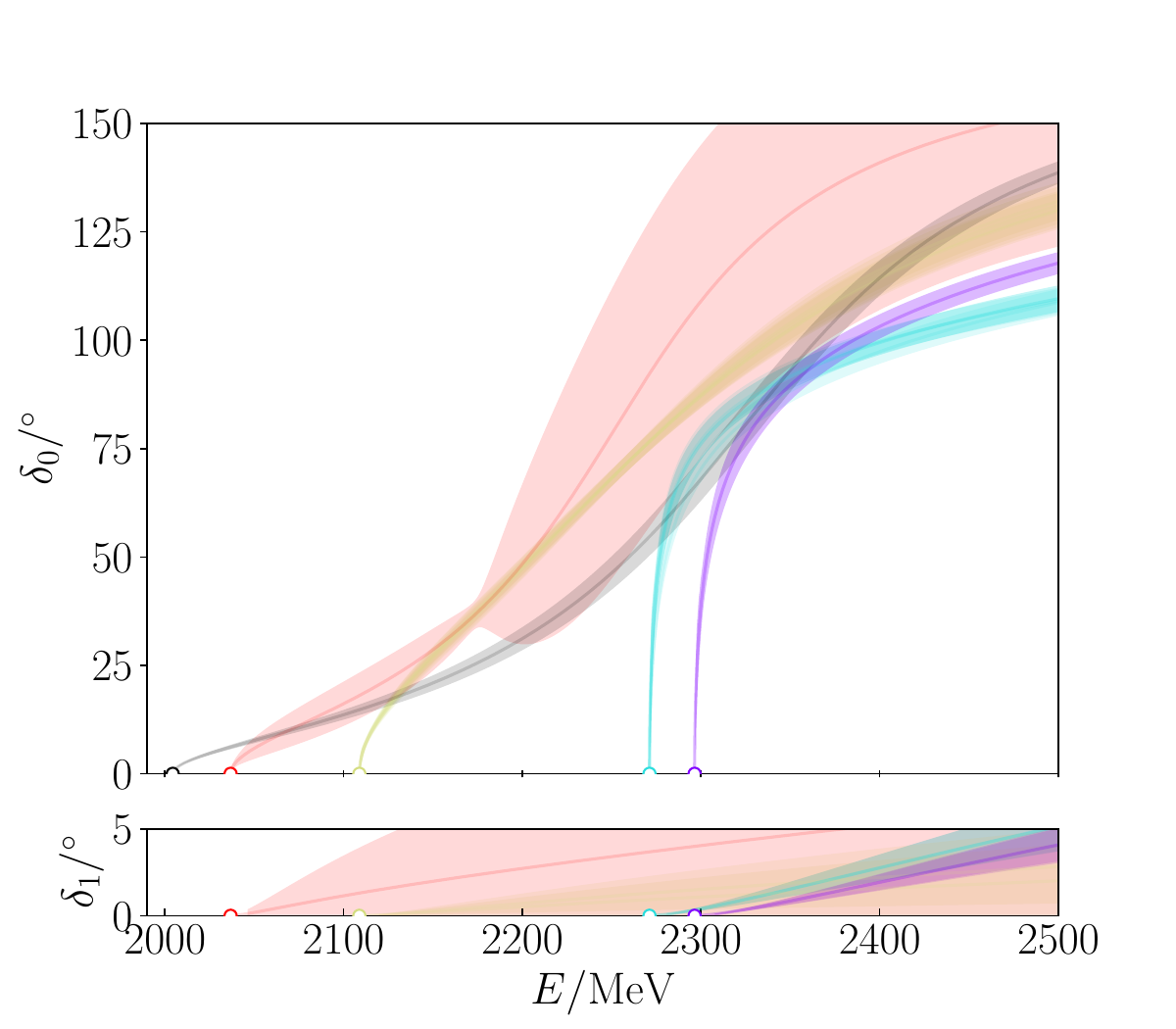}
\caption{Left: the finite-volume spectra at $m_{\pi} \approx 305$ MeV for irreps with leading partial waves being $S$- or $P$-wave. The black and gray points are the energy levels. The black lines, red dashed, and green dashed lines represent the non-interacting energies of the $D\pi$, $D^*\pi$, and the $D\pi\pi$, respectively. Only black points are used in the scattering analysis. The orange bands are the solutions of the L\"{u}scher's equations. Right: The $S$- and $P$-wave scattering phase shift $\delta_0$ (upper) and $\delta_1$ (lower). The red, yellow, cyan, and purple bands are results from $m_\pi \approx 133, 208, 305$ and $317$ MeV. The gray bands are reconstructed from the PDG value. The bands with the same color represent the results of different parametrizations at each pion mass while the width of each corresponds to the statistical error.}
\label{fig:spectra-Dpi-FH-1d2}
\end{figure*}

The spectra in each irrep is obtained by solving a generalized eigenvalue problem (GEVP)~\cite{Luescher1990}. The extracted spectra at $m_{\pi} \approx 305$ MeV are shown in Fig.~\ref{fig:spectra-Dpi-FH-1d2}. Non-negligible thermal pollution is found in the correlation functions. And the effect is addressed by generalizing the "weighting and shifting" method in Ref.~\cite{Dudek2012}. We refer to the detailed derivation, the solution, and the spectra at other pion masses in Ref.~\cite{supp}. There are two volumes at this pion mass. The black lines, red dashed, and green dashed lines represent the $D\pi$, $D^*\pi$, and the $D\pi\pi$ non-interacting levels, respectively. The black points are the energy levels obtained by fitting the eigenvalues of the correlation matrices. The orange bands are predictions from L\"{u}scher's equation (Eq.~\ref{eq:luscher}) which will be explained in the following.

For two-body scattering near the threshold, lower partial waves dominate. For irreps containing $P$-wave, a level much lower than the $D\pi$ threshold is found, corresponding to a vector $D^*$ bound state. For $[000] T_1^-$, the lowest $D\pi$ partial wave is the $P$-wave. The first excited states are not far from the non-interacting levels, indicating a weak interaction in the $P$-wave. For irreps containing $S$-wave as the leading partial wave, the $D\pi$ scattering energy levels are substantially lower than the corresponding non-interacting ones, indicating a strong attractive interaction.

\section{Scattering analysis}
\label{sec:scattering}
The finite volume spectra are related to the infinite-volume scattering phase shifts via L\"{u}scher's quantization condition, which can be written in the general form as~\cite{Briceno2014, Goeckeler2012},
\begin{equation}
\operatorname{det}[M(E, \vec{P} ; L)-\cot \delta(E)]=0,
\label{eq:luscher}
\end{equation}
where $M(E, \vec{P} ; L)$ contains the information of the finite volume spectra and $\delta$ is the phase shift. The scattering $t$-matrix is related to $\delta$ by $t_l = \frac{\sqrt{s}}{2} \frac{1}{k \cot\delta_l - ik}$ and can then be deduced from the lattice.

We cutoff the infinite-dimensional determinant in Eq.~\ref{eq:luscher} to $P$-wave. To extract both the $S$- and $P$-wave phase shifts, a correlated global fit is performed for energy levels below the lowest inelastic thresholds of the irreps with leading partial waves up to $P$-wave, shown in the left panel of Fig.~\ref{fig:spectra-Dpi-FH-1d2}. The number of energy levels is $17 \times 2 = 34$ for $m_\pi \approx 305$ and $208$ MeV, but is $17$ for $133$ and $317$ MeV, since there are two volumes for $m_\pi \approx 305$ and $208$ MeV.

To estimate the model dependencies, we apply many different parametrizations. The main results are reported by using the effective range expansion (ERE), $k^{2l+1} \cot \delta_{l}=\frac{1}{a_{l}}+\frac{1}{2} r_{l} k^2+P_2 k^4+\mathcal{O}(k^6)$, where for $l=0$, $a_0$ is the scattering length and $r_0$ is the effective range. The scattering momentum $k$ is related to the scattering energy by $E(\vec{k})=\sqrt{m_D^2+Z_D\vec{k}^2} + \sqrt{m_\pi^2+Z_\pi\vec{k}^2}$, where $Z_X$ is the square of the speed of light on the lattice. Due to lattice artifacts, $Z_X$ deviates from $1$. The values of $Z_{\pi}$, $Z_D$ and $Z_{D^*}$ are given in Ref.~\cite{supp}. For the ensembles except C48P14, we performed the scattering analysis using both the lattice values of $Z_X$ and the continuum value $1$. The difference is estimated as a systematic error. The fit results using the ERE are tabulated in Tab.~\ref{tab:ERE-Dpi-1d2}. The ensemble at $m_{\pi} \approx 133$ MeV with the Fermilab charm action is denoted as $133^*$. Although this procedure may not provide a precise measure of discretization error, the results remain unaffected, as the error estimated from this method is significantly smaller than the statistical error. Consequently, the total error is still dominated by statistical uncertainties, leaving the final results largely unchanged. For the coarsest ensemble, this difference is further highlighted by using two distinct charm actions, demonstrating that the variation lies well within the current statistical error. Other parametrizations and the details about the fitting can be found in Ref.~\cite{supp}. In general, little model dependence is found.

\begin{table}[htbp]
\caption{The fitting result for the $S$- and $P$-wave of $I=\frac{1}{2}$ $D\pi$ scattering using the ERE parametrization. $133^*$ denotes calculations performed on the same ensemble as $133$ but with a Fermilab charm action.}
\begin{tabular}{ccccc}
\toprule
$m_{\pi}/\text{MeV}$ & $a_0/\text{fm}$ & $r_0/\text{fm}$ & $a_1/\text{fm}^3$ & $r_1/\text{fm}^{-1}$ \\
\midrule
$317$ & $3.36(82)$ & $-0.734(60)$ & $0.0288(70)$ & $32(11)$ \\
$305$ & $3.9(1.2)$ & $-0.488(90)$ & $0.033(15)$ & $28(13)$ \\
$208$ & $0.663(60)$ & $-1.15(15)$ & $0.020(21)$ & $127(237)$ \\
$133$ & $0.33(16)$ & $-2.4(3.8)$ & $0.17(32)$ & $45(58)$ \\
$133^*$ & $0.26(14)$ & $-4.0(5.5)$ & $0.28(18)$ & $63(110)$ \\
\bottomrule
\end{tabular}
\label{tab:ERE-Dpi-1d2}
\end{table}

The energy spectra for any $L$ can then be reconstructed from the parameterized L\"{u}scher's formula, which are plotted as the orange bands in Fig.~\ref{fig:spectra-Dpi-FH-1d2}. The solutions from the L\"{u}scher's formula describe the data well.

The phase shifts for all the ensembles used in this work are plotted in the right panel of Fig.~\ref{fig:spectra-Dpi-FH-1d2}. There are $4$ bands for $m_{\pi} \approx 305, 317$ MeV and $2$ for $m_{\pi} \approx 208, 133$ MeV. The bands with the same color represent the results from various parametrizations at each pion mass with the width of the bands representing the statistical error. The differences among the bands are estimated as the systematic error due to different parameterizations. The gray curves are reconstructed from the PDG-averaged RBW parameters. The $S$-wave phase shift of all $m_{\pi}$'s rises from $0$ to $180$ degree, indicating a nontrivial structure. The $P$-wave phase shift $\delta_1$ is small compared to the $S$-wave in the elastic zone.

\section{Pole positions}
\label{sec:pole}
Observable structures normally appear as poles in the scattering amplitudes in the complex plane $s=E_{\text{CM}}^2$ plane. If there is an isolated pole near the real axis beyond the threshold, the scattering phase shifts, and the cross section would signal a sharp change close to the pole position.
The $t$-matrix near a pole behaves as $t \sim \frac{c^2}{s - s_0}$ with $s_0$ being the pole position. The mass $m$ and the width $\Gamma$ of the resonance are related the pole position by $\sqrt{s_0} = m - \frac{i}{2} \Gamma$.
To compare the results from different lattice groups, we focus on the mass of the structure subtracted by spin-averaged mass of $D$ and $D^{*}$, as was done in Ref.~\cite{Mohler2013}, $\sqrt{s_0^{\prime}} = \sqrt{s_0} - \frac{1}{4}(m_D + 3 m_{D^*})$, which we refer to as the spin-average-subtracted pole position.

\begin{figure}[htbp]
\centering
\includegraphics[width=\columnwidth]{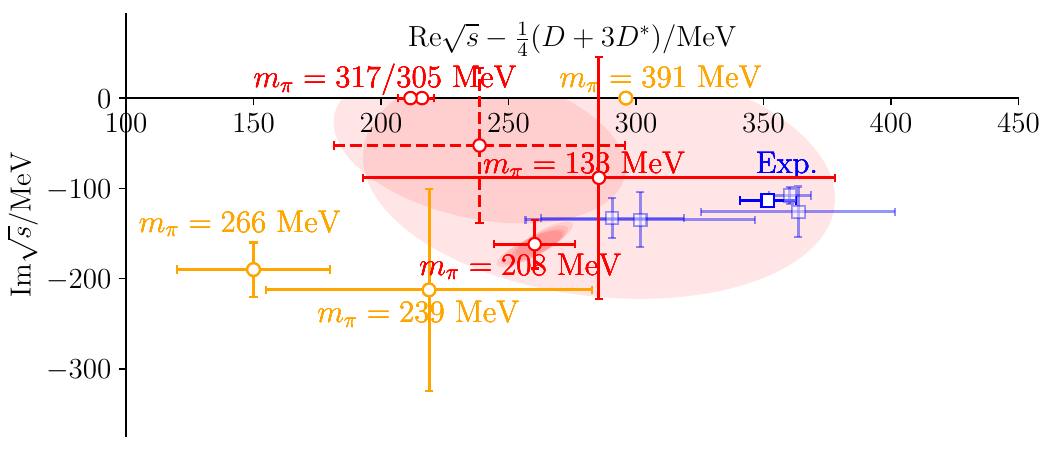}
\caption{The poles for $S$-wave of $I=\frac{1}{2}$ $D\pi$ scattering. The poles from this work are the covariance ellipses and the red points that wrap over them. The dashed red error bar indicates the result at $m_{\pi} \approx 133$ MeV obtained using the Fermilab action. The blue points are the experimental measurements~\cite{Abe2004, Aubert2009, Aaij2015, Aaij2015a}. The solid blue point is the PDG average~\cite{Workman2022}. The orange points are the results from other lattice calculations\cite{Mohler2013, Moir2016, Gayer2021}.}
\label{fig:pole-Dpi}
\end{figure}

For the $S$-wave of the $D\pi$ scattering, we found a virtual state for $m_{\pi} \approx 305$ and $317$ MeV, and a resonance pole for $m_{\pi} \approx 133$ and $208$ MeV, from all parametrizations. These poles are identified to the scalar $D_0^*$ found in the experiment. The spin-average-subtracted pole positions on the Riemann sheet are tabulated in Tab.~\ref{tab:pole} and shown as the covariance ellipses and the red error bars in Fig.~\ref{fig:pole-Dpi}. The poles from different parametrizations at the same $m_{\pi}$ are represented by the overlapping ellipses. The small deviations among different ellipses indicate that the parameterization dependence of the pole positions is small compared to the statistical error. The red error bars at $m_{\pi} \approx 208$ MeV cover the range of the pole positions from all parametrizations, and the discretization error is added in quadrature. The dashed red error bar indicates the result at $m_{\pi} \approx 133$ MeV obtained using the Fermilab action. The results at $m_{\pi} \approx 317$ MeV and $m_{\pi} \approx 305$ MeV are close to each other. Since the two results share similar pion masses but rather different lattice spacings, this indicates that the lattice discretization effect should be small for this lattice spacing. For the coarsest lattice C48P14 with $m_{\pi} \approx 133$ MeV, we have presented results with charm quark being either conventional Clover action or the Fermilab formulation to give an estimate of the lattice artifact. Although we cannot entirely rule out the possibility that the effects of differing pion masses and lattice spacings may cancel each other out, we still expect the discretization effect to be minimal due to the similarity in pion masses, at least for the current statistical error.
In the figure, we have also presented the spin-average-subtracted pole positions from other lattice calculations which were done at $m_{\pi} \approx 266$, $239$ and $391$ MeV.~\cite{Mohler2013, Moir2016, Gayer2021}. The plot also presents the pole positions of the experimental results that describe $D_0^*$ by RBW. These are the blue points. The solid blue data are from the PDG average value~\cite{Workman2022} and the transparent ones are the experimental measurements~\cite{Abe2004, Aubert2009, Aaij2015, Aaij2015a} it uses.

\begin{table}[htbp]
\caption{The pole positions for $S$-wave of $I=\frac{1}{2}$ $D\pi$ scattering. The errors account for the statistical error, the variance across multiple parametrizations, and the estimate of the lattice discretization effects. $133^*$ denotes calculations performed on the same ensemble as $133$ but with a Fermilab charm action.}
\addtolength{\tabcolsep}{6pt}
\begin{tabular}{ccc}
\toprule
$m_{\pi}/\text{MeV}$ & $\operatorname{Re} \sqrt{s} - \frac{1}{4} (D + 3 D^*) /\text{MeV}$ & $\operatorname{Im} \sqrt{s} /\text{MeV}$ \\
\midrule
$317$ & $211.6(4.9)$ & $0$ \\
$305$ & $216.1(4.8) $ & $0.00(11)$ \\
$208$ & $260(16)$ & $162(27)$ \\
$133$ & $285(93)$ & $88(134)$ \\
$133^*$ & $239(57)$ & $53(86)$ \\
\bottomrule
\end{tabular}
\addtolength{\tabcolsep}{-6pt}
\label{tab:pole}
\end{table}

Combining the results from the previous lattice calculations~\cite{Mohler2013, Moir2016, Gayer2021}, an interesting running behavior of the $D_0^*$ pole as a function of $m_{\pi}$ emerges: The pole starts at $m_{\pi} \approx 391$ MeV on the real axis as a bound state and becomes lighter as $m_{\pi}$ decreases. At some point in $317$ MeV $\lesssim m_{\pi} \lesssim 391$ MeV, the pole penetrates the Riemann sheet and passes to the unphysical sheet where it becomes a virtual state. As $m_{\pi}$ further decreases, the pole evolves to the complex plane at $305$ MeV $\lesssim m_{\pi} \lesssim 266$ MeV and becomes a resonance, and the pole stops moving leftwards but turns rightwards and begins to move on a complicated trajectory. The pole at the physical pion mass is consistent with the experiments. However, due to the large uncertainty, no conclusive statement should be made. It is also interesting to note that the trajectory of the $D_0^*$ poles is qualitatively similar to the predictions from the chiral perturbation theory~\cite{Guo2009, Guo2015}. See Ref.~\cite{Hanhart2014, Matuschek2021} for a general discussion.

\section{Chiral interpolations of $a_0^{-1}$}
\label{sec:interpolations}
In this section, we interpolate the $S$-wave scattering length to the physical pion mass $m_{\pi}^{\text{phys}} = 135$~MeV. The systematic error of the scattering length from the model dependence is included by wrapping the spread of results of $\lim_{k \to 0} k \cot \delta_0$ from all parametrizations. The discretization error is also included. We use a polynomial form to describe the chiral behavior,
\begin{equation}
a_0^{-1}(m_{\pi}) = c_0 + c_1 m_{\pi}^2 + c_2 m_{\pi}^4.
\label{eq:chiral}
\end{equation}
Similar extrapolations were done in Ref.~\cite{Lyu2023}. The fit uses determinations of  $a_0^{-1}$ from simulations with the Clover charm action to maintain consistency in the action; hence, the $133^*$ results are not included in the extrapolation. The fitting results are $c_0=4.66(95) \ \text{fm}^{-1}$, $c_1=-94(29) \ \text{MeV}^{-2} \cdot \text{fm}^{-1}$, $c_2=502(203) \ \text{MeV}^{-4} \cdot \text{fm}^{-1}$ with $\chi^2/\text{d.o.f} = 1.02$. The inverse scattering length at the physical point is then
\begin{equation}
a_0^{-1}(m_{\pi}^{\text{phys}}) = 3.12(50) \ \text{fm}^{-1}.
\label{eq:a0}
\end{equation}
Adding to Eq.~\ref{eq:chiral} an $a^2$-term yields consistent results, indicating that the data points reasonably incorporate the estimation of the discretization error. Due to the large statistical error at the physical pion mass, the chiral extrapolation/interpolation is essentially constrained by data points at $208$ MeV and $305/317$ MeV. However, two observations most likely support the reliability of the extrapolation/interpolation. First, data points from previous lattice calculations align closely with the chiral curves obtained in this study. Second, the value of $a_0^{-1}(m_{\pi}^{\text{phys}})$ is consistent with the direct calculation at the physical ensemble. These factors suggest that the results are dependable despite the aforementioned limitations.

\begin{figure}[htbp]
\centering
\begin{tikzpicture}
\pgftext{
\includegraphics[width=\columnwidth]{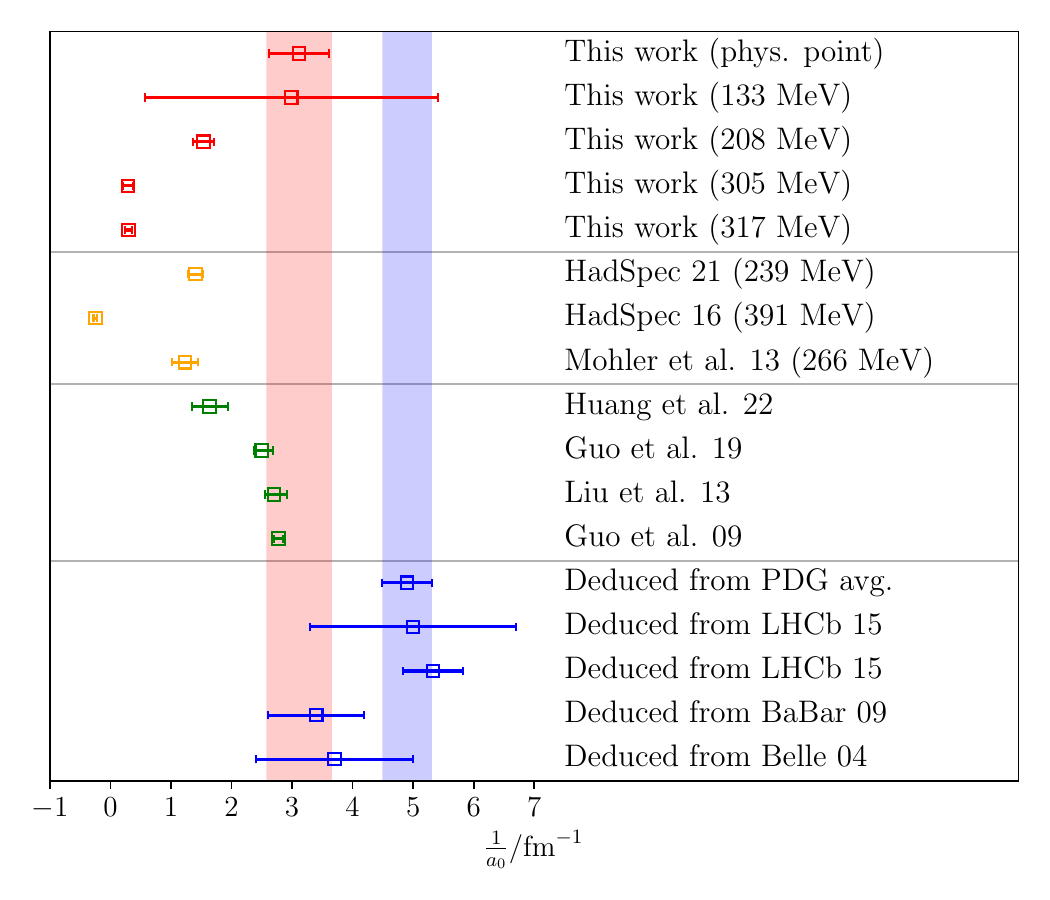}
}
\node[font = \small, anchor = west] at (2.9, 1.46) {\cite{Gayer2021}};
\node[font = \small, anchor = west] at (2.9, 1.46-1*0.3654545455) {\cite{Moir2016}};
\node[font = \small, anchor = west] at (3.4, 1.46-2*0.3654545455) {\cite{Mohler2013}};
\node[font = \small, anchor = west] at (2.1, 1.46-3*0.3654545455) {\cite{Huang2022}};
\node[font = \small, anchor = west] at (1.9, 1.46-4*0.3654545455) {\cite{Guo2019}};
\node[font = \small, anchor = west] at (1.8, 1.46-5*0.3654545455) {\cite{Liu2013}};
\node[font = \small, anchor = west] at (1.9, 1.46-6*0.3654545455) {\cite{Guo2009}};
\node[font = \small, anchor = west] at (3.1, 1.46-7*0.3654545455) {\cite{Workman2022}};
\node[font = \small, anchor = west] at (3.0, 1.46-8*0.3654545455) {\cite{Aaij2015}};
\node[font = \small, anchor = west] at (3.0, 1.46-9*0.3654545455) {\cite{Aaij2015a}};
\node[font = \small, anchor = west] at (3.1, 1.46-10*0.3654545455) {\cite{Aubert2009}};
\node[font = \small, anchor = west] at (2.9, -2.56) {\cite{Abe2004}};
\end{tikzpicture}
\caption{The comparison of $\frac{1}{a_0}$ from this work (red), the previous lattice results~\cite{Mohler2013,Moir2016, Gayer2021} (orange), the experiments~\cite{Workman2022, Abe2004, Aubert2009, Aaij2015, Aaij2015a} (blue), and the phenomenological models~\cite{Huang2022, Guo2019, Liu2013, Guo2009} (green). The experimental values are obtained by analytically continuing the RBW amplitudes from the measured parameters.}
\label{fig:comparison-Dpi-1d2}
\end{figure}

With no direct experimental report of the scattering length available, we infer it by analytically continuing the RBW amplitudes and taking the threshold limit. The result from~\cite{Mohler2013} is also presented similarly. The scattering lengths in this work together with previous lattice results~\cite{Mohler2013, Moir2016, Gayer2021}, the deduced PDG average~\cite{Workman2022}, the experimental measurements~\cite{Abe2004, Aubert2009, Aaij2015, Aaij2015a}, and the phenomenological models~\cite{Huang2022, Guo2019, Liu2013, Guo2009} are shown in Fig.~\ref{fig:comparison-Dpi-1d2}. Limited by only one or two values in previous lattice calculations, the dependence on the pion mass is hardly recognized, which leads to poor knowledge of the discrepancy with experiments. Given the red points in this work, a clearly visible $m_{\pi}$ dependency is identified and the discrepancy with the PDG value is also greatly alleviated. In fact, our result is consistent with the BaBar and Belle experiments within the range of error. The result is also consistent with those in the phenomenological models in Ref.~\cite{Guo2019, Liu2013, Guo2009}.

\section{Conclusions}
\label{sec:conclusions}
Using a series of $N_{\text{f}}=2+1$ Wilson-Clover configurations, the $S$- and $P$-wave $I=\frac{1}{2}$ $D\pi$ scattering phases are studied at $m_\pi \approx 133, 208, 305$ and $317$ MeV within L\"uscher's formalism. Various single-meson and two-meson interpolating operators are constructed, which leads to stable excited states energy spectra from variational analysis of the correlation matrices. 
A strong attractive interaction in irreps that contain $S$-wave is observed while that of $P$-wave is small. The scattering phase shifts are extracted by solving the L\"uscher's equation. 
The $D_0^*$ found on our configurations are virtual states at $m_\pi \approx 305, 317$ MeV and are resonance states at $m_\pi \approx 133, 208$ MeV. Together with the results from previous works, we arrive at the following picture for the movement of the $D_0^*$ pole when $m_\pi$ varies: $D_0^*$ pole is a bound state at $m_{\pi} \gtrsim 391$ MeV, it then evolves into a virtual state somewhere in $317$ MeV $\lesssim m_{\pi} \lesssim 391$ MeV and finally becomes a resonance at $305$ MeV $\lesssim m_{\pi} \lesssim 266$ MeV. The pole moves to the left at first, but then turns around and moves to the right and shows a convoluted behavior. This trajectory of the $D_0^*$ pole as a function of pion mass provides an essential clue for phenomenological models.

The scattering lengths are interpolated to the physical pion mass, which is then compared to the deduced experimental measurements. There is still tension between our result and the PDG-averaged result. Future calculations performed at finer lattice spacings and more pion masses would be valuable to further clarify this. Furthermore, the results presented in this Letter on the $D\pi$ scattering amplitude near the threshold could be utilized for future $DD\pi$ three-body studies.

Finally, we found only one pole near the threshold region. 
The missing of the two-pole structure is likely due to the ignorance of the coupled-channel effects. To solve the puzzle of the mass ordering of $D_0^*$ and $D_{s0}^*$,  a similar investigation in $DK$ scattering is underway.

\begin{acknowledgments}
\

H.~Yan is grateful to L.~An, F.~Guo, C.~Hanhart, B.~H\"{o}rz, M.~Garofalo, M.~Mai, S.~Prelov\v{s}ek, C.~Urbach, and J.~Wu for valuable discussions, and extends appreciation to members of CLQCD, including Y.~Chen, X.~Feng, P.~Sun, for collaborations and comments. Software \verb|Chroma|~\cite{Edwards2005} and \verb|QUDA|~\cite{Clark2010, Babich2011, Clark2016} are used to generate the configurations and solve the perambulators. H.~Yan, C.~Liu, L.~Liu, and Y.~Meng acknowledge support from NSFC under Grant No. 12293060, 12293061, 12293063, 12175279, 12305094, 11935017. L.~Liu thanks support from the Strategic Priority Research Program of the Chinese Academy of Sciences with Grant No. XDB34030301 and Guangdong Major Project of Basic and Applied Basic Research No. 2020B0301030008. This work is also supported by the Deutsche Forschungsgemeinschaft (DFG, German Research Foundation) through the Sino-German Collaborative Research Center CRC110 “Symmetries and the Emergence of Structure in QCD” (DFG Project ID 196253076 - TRR 110). Part of the simulations were performed on the High-performance Computing Platform of Peking University, the Southern Nuclear Science Computing Center (SNSC), the Dongjiang Yuan Intelligent Computing Center, and the SongShan supercomputer at Zhengzhou National Supercomputing Center.
\end{acknowledgments}

\bibliographystyle{apsrev4-2}
\bibliography{paper}

\appendix
\onecolumngrid
\clearpage

\section*{Supplemental Material}
\subsection{Table of operators}
\label{sec:supp:operators}
Since the physical process happens on a cubic spacetime, the rotational symmetries are broken into the cubic group, and its subgroups which are further reduced by the Lorentz boost. The finite-volume spectra are also subduced into the irreducible representations (irreps) of the little groups. We consider reference frames with a center of mass (CM) three-momenta $\vec{P}$= [0,0,0], [0,0,1], [0,1,1], [1,1,1], [0,0,2] in the unit of $\frac{2\pi}{L}$.

For the construction of the single-hadron operators, the helicity strategy proposed by the Hadron Spectrum Collaboration~\cite{Dudek2010, Thomas2012} is applied. For the construction of the two-hadron operators, we propose a generalized version of the projection formula put forward by 
Prelovsek \textit{et. al}~\cite{Prelovsek2017},
\begin{equation}
O_{\Gamma, \mu}^{[|k|]}(\vec{P}) = \sum_{R \in G} T_{\mu, \mu}^{\Gamma}(R) R O_1(\vec{k}) O_2(\vec{P}-\vec{k}) R^{\dagger},
\end{equation}
where $\vec{k}$ designates the relative momentum between the two single hadrons. $T_{\mu, \mu}^{\Gamma}$ is the diagonal matrix elements $\mu,\mu$ of the irrep $\Gamma$. $O_{1,2}$ are single-hadron operators of the form $\bar{\psi} \Gamma \psi$ with $\bar{\psi}$ and $\psi$ being light or charm quarks and the gamma matrices transform as scalar and vector.

We consider reference frames with a center of mass (CM) three-momenta $\vec{P}$= [0,0,0], [0,0,1], [0,1,1], [1,1,1], [0,0,2] in the unit of $\frac{2\pi}{L}$, which corresponds to the cubic group $O_h$ and the little groups $C_{4v}, C_{2v}, C_{3v}, C_{4v}$, respectively. We consider all irreps with leading partial waves up to $P$-wave for $D\pi$ scattering. Note that since $m_D \neq m_{\pi}$, $S$- and $P$-wave can mix in moving frames where parity is no longer a good quantum number. For the $D\pi$ system, up to $P$-wave, the irreps $A_1$ of $C_{4v}, C_{2v}, C_{3v}$ contain both $S$ and $P$-wave; irreps $T_1^-$ of $O_h$, $E_2$ of $C_{4v}, C_{3v}$ and $B_1, B_2$ of $C_{2v}$ house only $P$-wave; $A_1^+$ of $O_h$ houses only $S$-wave. 

It is useful to do a linear transformation of the operator set if possible. We perform the transformation in the irreps in the CM frame to let the operators transform in a certain partial wave, as constructed by the partial-wave method in Ref.~\cite{Prelovsek2017}. In this way, the operators mainly couple to the low partial waves we are concerned.

The operators need to be further projected to isospin $I=\frac{1}{2}$,
\begin{equation}
    O_{D^{(*)}\pi}^{I=\frac{1}{2}} = 2 D^0\pi^+ - \sqrt{2} D^{(*)+}\pi^0,
\label{eq:isospin}
\end{equation}
The single-hadron operators appeared with different electric charges had the same flavor structure and differed only in isospin.

Each operator is then a linear combination of quark bilinear or the product of two bilinears. The operators are represented as
\begin{equation}
\begin{cases}
    O_{\text{one}}(\vec{P}) = \sum_{i} \eta_i D_{\mu_i}(\vec{P}), \\
    O_{\text{two}}(\vec{P}) = \sum_{i} \eta_i D_{\mu_i}(\vec{p}_i) \pi(\vec{P}-\vec{p}_i),
\end{cases}
\end{equation}
and are uniquely identified by the parameters $\eta_i$, $\mu_i$ and $\vec{p}_i$ for one- and two-meson operators, respectively. $\mu_i \in \{ 0, 5, x, y, z, 5x, 5y, 5z \}$ is the Cartesian gamma matrices and corresponds to $\gamma_0, \gamma_5, \gamma_x, \gamma_y, \gamma_z, \gamma_5\gamma_x, \gamma_5\gamma_y, \gamma_5\gamma_z$. $\vec{p}_i$ is the relative momenta. For convenience, the parameters are denoted by
\begin{equation}
\eta_{\mu_i}^{\alpha_i},
\end{equation}
where $\alpha_i$ one-to-one corresponds to the momentum vector. One direction notation in $\alpha_i$ means one unit of momentum in that direction. For example, $\alpha = yz \to \vec{p} = [011]$, $\alpha = -2x \to \vec{p} = [-200]$, \textit{etc}. Operators for isospin $I=\frac{1}{2}$ for $\vec{P} = [000]$ is shown in Tab.~\ref{tab:operators000}; $\vec{P} = [001]$ is shown in Tab.~\ref{tab:operators001}; $\vec{P} = [011]$ is shown in Tab.~\ref{tab:operators011}; $\vec{P} = [111]$ is shown in Tab.~\ref{tab:operators111}; $\vec{P} = [002]$ is shown in Tab.~\ref{tab:operators002}.

We have also tested that adding to the operator set the covariant derivatives makes the resulting energy levels plateau a little earlier, but the signals become significantly worse. As a result, the operators with covariant derivatives are not applied in this work.

\begin{table*}[htbp]
\caption{Operators for $\vec{P} = [000]$ used to interpolate the $D\pi$ system. Each irrep contains both one-meson type and two-meson type operators. The symbols $\eta_{\mu_i}^{\alpha_i}$ are defined in the context. Each operator consists of a linear combination of one- or two-hadron quark bilinears. $\eta$ denotes the coefficients of certain operators. $\mu$ indicates the Cartesian gamma matrices. $\alpha$ denotes the momentum vector of the single hadron or the relative momentum vector of the two hadrons.}
\begin{tabular}{ccc}
\toprule
irrep & type & operator \\
\midrule
\multirow{4}{*}{$A_1^+$} & \multirow{1}{*}{one} & $(+1)^{0}_{0}$ \\
\cmidrule(lr){2-3}
& \multirow{3}{*}{two} & $(+1)^{0}_{5}$ \\
& & $(+1)^{z}_{5}, (+1)^{y}_{5}, (+1)^{x}_{5}, (+1)^{-z}_{5}, (+1)^{-y}_{5}, (+1)^{-x}_{5}$ \\
& & $(+1)^{yz}_{5}, (+1)^{xz}_{5}, (+1)^{xy}_{5}, (+1)^{-y,z}_{5}, (+1)^{-x,z}_{5}, (+1)^{-x,y}_{5}, (+1)^{y,-z}_{5}, (+1)^{x,-z}_{5}, (+1)^{x,-y}_{5}, (+1)^{-y,-z}_{5}, (+1)^{-x,-z}_{5}, (+1)^{-x,-y}_{5}$ \\
\midrule
\multirow{4}{*}{$T_1^-$} & \multirow{1}{*}{one} & $(+1)^{0}_{z}$ \\
\cmidrule(lr){2-3}
& \multirow{3}{*}{two} & $(+1)^{z}_{5}, (-1)^{-z}_{5}$ \\
& & $(+1)^{xz}_{5}, (+1)^{-x,z}_{5}, (+1)^{yz}_{5}, (+1)^{-y,z}_{5}, (-1)^{x,-z}_{5}, (-1)^{-x,-z}_{5}, (-1)^{y,-z}_{5}, (-1)^{-y,-z}_{5}$ \\
& & $(+1)^{y}_{x}, (-1)^{-y}_{x}, (+1)^{-x}_{y}, (-1)^{x}_{y}$ \\
\midrule
\multirow{4}{*}{$T_1^+$} & \multirow{1}{*}{one} & $(+1)^{0}_{5z}$ \\
\cmidrule(lr){2-3}
& \multirow{3}{*}{two} & $(+1)^{0}_{z}$ \\
& & $(+1)^{z}_{z}, (+1)^{y}_{z}, (+1)^{x}_{z}, (+1)^{-z}_{z}, (+1)^{-y}_{z}, (+1)^{-x}_{z}$ \\
& & $(-2)^{z}_{z}, (+1)^{y}_{z}, (+1)^{x}_{z}, (-2)^{-z}_{z}, (+1)^{-y}_{z}, (+1)^{-x}_{z}$ \\
\bottomrule
\end{tabular}
\label{tab:operators000}
\end{table*}

\begin{table}[htbp]
\caption{Operators for $\vec{P} = [001]$ used to interpolate the $D\pi$ system. Notations as in Tab.~\ref{tab:operators000}.}
\begin{tabular}{ccc}
\toprule
irrep & type & operator \\
\midrule
\multirow{8}{*}{$A_1$} & \multirow{2}{*}{one} & $(+1)^{z}_{0}$ \\
& & $(+1)^{z}_{z}$ \\
\cmidrule(lr){2-3}
& \multirow{6}{*}{two} & $(+1)^{0}_{5}$ \\
& & $(+1)^{z}_{5}$ \\
& & $(+1)^{-x}_{5}, (+1)^{x}_{5}, (+1)^{-y}_{5}, (+1)^{y}_{5}$ \\
& & $(+1)^{xz}_{5}, (+1)^{-x,z}_{5}, (+1)^{yz}_{5}, (+1)^{-y,z}_{5}$ \\
& & $(+1)^{-z}_{5}$ \\
& & $(+1)^{2z}_{5}$ \\
\midrule
\multirow{6}{*}{$E_2$} & \multirow{2}{*}{one} & $(+1)^{z}_{y}$ \\
& & $(+1)^{z}_{5x}$ \\
\cmidrule(lr){2-3}
& \multirow{4}{*}{two} & $(+1)^{y}_{5}, (-1)^{-y}_{5}$ \\
& & $(+1)^{yz}_{5}, (-1)^{-y,z}_{5}$ \\
& & $(+1)^{0}_{x}$ \\
& & $(+1)^{z}_{x}$ \\
\bottomrule
\end{tabular}
\label{tab:operators001}
\end{table}

\begin{table}[htbp]
\caption{Operators for $\vec{P} = [011]$ used to interpolate the $D\pi$ system. Notations as in Tab.~\ref{tab:operators000}.}
\begin{tabular}{ccc}
\toprule
irrep & type & operator \\
\midrule
\multirow{8}{*}{$A_1$} & \multirow{2}{*}{one} & $(+1)^{yz}_{0}$ \\
& & $(+1)^{yz}_{y}, (+1)^{yz}_{z}$ \\
\cmidrule(lr){2-3}
& \multirow{6}{*}{two} & $(+1)^{0}_{5}$ \\
& & $(+1)^{yz}_{5}$ \\
& & $(+1)^{y}_{5}, (+1)^{z}_{5}$ \\
& & $(+1)^{xy}_{5}, (+1)^{xz}_{5}, (+1)^{-x,y}_{5}, (+1)^{-x,z}_{5}$ \\
& & $(+1)^{xyz}_{5}, (+1)^{-x,yz}_{5}$ \\
& & $(+1)^{y}_{x}, (-1)^{z}_{x}$ \\
\midrule
\multirow{5}{*}{$B_1$} & \multirow{2}{*}{one} & $(+1)^{yz}_{y}, (-1)^{yz}_{z}$ \\
& & $(+1)^{yz}_{5x}$ \\
\cmidrule(lr){2-3}
& \multirow{3}{*}{two} & $(+1)^{y}_{5}, (-1)^{z}_{5}$ \\
& & $(+1)^{xy}_{5}, (-1)^{xz}_{5}, (+1)^{-x,y}_{5}, (-1)^{-x,z}_{5}$ \\
& & $(+1)^{y}_{x}, (+1)^{z}_{x}$ \\
\midrule
\multirow{9}{*}{$B_2$} & \multirow{2}{*}{one} & $(+1)^{yz}_{x}$ \\
& & $(+1)^{yz}_{5y}, (-1)^{yz}_{5z}$ \\
\cmidrule(lr){2-3}
& \multirow{7}{*}{two} & $(+1)^{xy}_{5}, (+1)^{xz}_{5}, (-1)^{-x,y}_{5}, (-1)^{-x,z}_{5}$ \\
& & $(+1)^{x}_{5}, (-1)^{-x}_{5}$ \\
& & $(+1)^{xyz}_{5}, (-1)^{-x,yz}_{5}$ \\
& & $(+1)^{y}_{y}, (-1)^{z}_{z}$ \\
& & $(+1)^{y}_{z}, (-1)^{z}_{y}$ \\
& & $(+1)^{0}_{y}, (-1)^{0}_{z}$ \\
& & $(+1)^{yz}_{y}, (-1)^{yz}_{z}$ \\
\bottomrule
\end{tabular}
\label{tab:operators011}
\end{table}

\begin{table}[htbp]
\caption{Operators for $\vec{P} = [111]$ used to interpolate the $D\pi$ system. Notations as in Tab.~\ref{tab:operators000}.}
\begin{tabular}{ccc}
\toprule
irrep & type & operator \\
\midrule
\multirow{8}{*}{$A_1$} & \multirow{2}{*}{one} & $(+1)^{xyz}_{0}$ \\
& & $(+1)^{xyz}_{x}, (+1)^{xyz}_{y}, (+1)^{xyz}_{z}$ \\
\cmidrule(lr){2-3}
& \multirow{6}{*}{two} & $(+1)^{0}_{5}$ \\
& & $(+1)^{xyz}_{5}$ \\
& & $(+1)^{x}_{5}, (+1)^{y}_{5}, (+1)^{z}_{5}$ \\
& & $(+1)^{yz}_{5}, (+1)^{xz}_{5}, (+1)^{xy}_{5}$ \\
& & $(+1)^{x}_{y}, (-1)^{x}_{z}, (+1)^{y}_{z}, (-1)^{y}_{x}, (+1)^{z}_{x}, (-1)^{z}_{y}$ \\
& & $(+1)^{xy}_{x}, (-1)^{xy}_{y}, (+1)^{yz}_{y}, (-1)^{yz}_{z}, (+1)^{xz}_{z}, (-1)^{xz}_{x}$ \\
\midrule
\multirow{8}{*}{$E_2$} & \multirow{2}{*}{one} & $(+1)^{xyz}_{x}, (-1)^{xyz}_{y}$ \\
& & $(+1)^{xyz}_{5x}, (+1)^{xyz}_{5y}, (-2)^{xyz}_{5z}$ \\
\cmidrule(lr){2-3}
& \multirow{6}{*}{two} & $(+1)^{x}_{5}, (-1)^{y}_{5}$ \\
& & $(+1)^{yz}_{5}, (-1)^{xz}_{5}$ \\
& & $(+1)^{0}_{x}, (+1)^{0}_{y}, (-2)^{0}_{z}$ \\
& & $(+1)^{xyz}_{x}, (+1)^{xyz}_{y}, (-2)^{xyz}_{z}$ \\
& & $(+1)^{x}_{x}, (+1)^{y}_{y}, (-2)^{z}_{z}$ \\
& & $(+1)^{yz}_{x}, (+1)^{xz}_{y}, (-2)^{xy}_{z}$ \\
\bottomrule
\end{tabular}
\label{tab:operators111}
\end{table}

\begin{table}[htbp]
\caption{Operators for $\vec{P} = [002]$ used to interpolate the $D\pi$ system. Notations as in Tab.~\ref{tab:operators000}.}
\begin{tabular}{ccc}
\toprule
irrep & type & operator \\
\midrule
\multirow{6}{*}{$A_1$} & \multirow{2}{*}{one} & $(+1)^{2z}_{0}$ \\
& & $(+1)^{2z}_{z}$ \\
\cmidrule(lr){2-3}
& \multirow{4}{*}{two} & $(+1)^{0}_{5}$ \\
& & $(+1)^{2z}_{5}$ \\
& & $(+1)^{z}_{5}$ \\
& & $(+1)^{xz}_{5}, (+1)^{-x,z}_{5}, (+1)^{yz}_{5}, (+1)^{-y,z}_{5}$ \\
\midrule
\multirow{6}{*}{$E_2$} & \multirow{2}{*}{one} & $(+1)^{2z}_{y}$ \\
& & $(+1)^{2z}_{5x}$ \\
\cmidrule(lr){2-3}
& \multirow{4}{*}{two} & $(+1)^{yz}_{5}, (-1)^{-y,z}_{5}$ \\
& & $(+1)^{0}_{x}$ \\
& & $(+1)^{z}_{x}$ \\
& & $(+1)^{yz}_{x}, (+1)^{-y,z}_{x}$ \\
\bottomrule
\end{tabular}
\label{tab:operators002}
\end{table}

We have developed a Mathematica package \verb|OpTion| (Operator construcTion) that can construct general $N$-hadron operators. By setting the type of hadrons, the desired quantum numbers, the number of derivatives, and the maximum relative momentum, the operator set can be obtained by one line of straightforward commands. The formalism of the operator construction will be described in detail by an independent publication~\cite{Yan2024}.

\subsection{Correlator analysis and thermal pollutions}
\label{sec:supp:thermal}
The distillation method~\cite{Peardon2009} is used to reduce the cost of generating propagators, which are calculated only in the subspace spanned by the $N_v$ lowest eigenvectors of the covariant Laplacian operator. The quark fields are smeared by the distillation operator such that the high-lying modes are suppressed. And $C_{ij}(t)$'s are then estimated using relevant perambulators. The $N_v$ applied in this work is $100, 200$ for ensembles with volume $32^3$ and $48^3$, respectively. It has been checked that the spectra are insensitive to $N_v$ and the error would slightly increase with smaller values of $N_v$.

To extract the finite-volume spectra, we solve the generalized eigenvalue problem (GEVP)~\cite{Luescher1990} of the correlation matrices for each irrep,
\begin{equation}
C(t) v_n(t, t_0) = \lambda_n(t, t_0) C(t_0) v_n(t, t_0),
\end{equation}
where the reference time $t_0 = 20$ for H48P32 and $t_0 = 10$ for other configurations. Changing $t_0$ hardly affects the resulting spectra. The energy levels of the $n^{\text{th}}$ excited states are extracted by fitting the eigenvalues $\lambda_n(t, t_0)$ 
\begin{equation}
\lambda_n(t, t_0) = (1-A_n) \operatorname{e}^{-E_n(t-t_0)} + A_n \operatorname{e}^{-E_n^{\prime}(t-t_0)}.
\end{equation}

It is found that the effective mass of some eigenvalues does not plateau at large time, but keeps rising or dropping. We found such behavior is due to the non-negligible thermal pollution which can be explained by considering the decomposition of the correlation matrix. We attribute the effect to the elements whose source and sink are both $D\pi$-like two-body operators with the same momentum structure:
\begin{equation}
\begin{aligned}
    &C_{[D\pi]_i,[D\pi]_j}(t) = \langle D_{\vec{k}}^{-}(t) \pi_{\vec{P} - \vec{k}}^{-}(t) D_{\vec{k}}^{+}(0) \pi_{\vec{P} - \vec{k}}^{+}(0) \rangle_T \\
    &\supset \frac{1}{Z_T} \left[ e^{-E_\pi(\vec{P} - \vec{k}) T} \langle \pi_{\vec{P} - \vec{k}}^{-} | D_{\vec{k}}^{-}(t) \pi_{\vec{P} - \vec{k}}^{-}(t) | D_{\vec{k}}^{+} \rangle \langle D_{\vec{k}}^{+} | D_{\vec{k}}^{+}(0) \pi_{\vec{P} - \vec{k}}^{+}(0) | \pi_{\vec{P} - \vec{k}}^{-} \rangle \right. \\
    &+ \left. e^{-E_D(\vec{k}) T} \langle D_{\vec{k}}^{-} | D_{\vec{k}}^{-}(t) \pi_{\vec{P} - \vec{k}}^{-}(t) | \pi_{\vec{P} - \vec{k}}^{+} \rangle \langle \pi_{\vec{P} - \vec{k}}^{+} | D_{\vec{k}}^{+}(0) \pi_{\vec{P} - \vec{k}}^{+}(0) | D_{\vec{k}}^{-} \rangle \right] \\
    &= \frac{1}{Z_T} |z_{\vec{k}}^D|^2 |z_{\vec{P} - \vec{k}}^{\pi}|^2  \left[ e^{-E_\pi(\vec{P} - \vec{k}) T -(E_D(\vec{k}) - E_\pi(\vec{P} - \vec{k})) t} + e^{-E_D(\vec{k}) T -(E_\pi(\vec{P} - \vec{k}) - E_D(\vec{k})) t} \right] \\
    &\approx \frac{1}{Z_T} |z_{\vec{k}}^D|^2 |z_{\vec{P} - \vec{k}}^{\pi}|^2 e^{-E_\pi(\vec{P} - \vec{k}) T -(E_D(\vec{k}) - E_\pi(\vec{P} - \vec{k})) t},
\end{aligned}
\end{equation}
where $z_{\vec{k}}^X \equiv \langle X_{\vec{k}}^{+} | X_{\vec{k}}^{+} | \Omega \rangle$. In the last step, the second thermal state is discarded since it is sub-leading, compared to the first term.

We generalize the "weighting and shifting" method as described in Ref.~\cite{Dudek2012} to the $D\pi$ system to eliminate the leading thermal state pollution. The leading thermal pollution can be removed by constructing the weighted-shifted correlation matrices,
\begin{equation}
\begin{aligned}
    \tilde{C}_{ij}(t) &= e^{(E_D(\vec{k}) - E_\pi(\vec{P} - \vec{k})) t} C_{ij}(t) - e^{(E_D(\vec{k}) - E_\pi(\vec{P} - \vec{k})) (t+1)} C_{ij}(t+1).
\end{aligned}
\end{equation}
Most elements have no such thermal state but the procedure does not affect their behavior under $t$. The spectra should be shifted up by $(E_D(\vec{k}) - E_\pi(\vec{P} - \vec{k}))$ to compensate for the weighting procedure. The weighted-shifted correlation matrices $\tilde{C}_{ij}(t)$ can then be sent through the usual GEVP process, yielding the right energy levels. By doing correlated fits of these eigenvalues, towers of energy spectra in different irreps can be obtained.

\subsection{Finite-volume spectra}
\label{sec:supp:spectra}
\begin{figure*}[htbp]
\includegraphics[width=\textwidth]{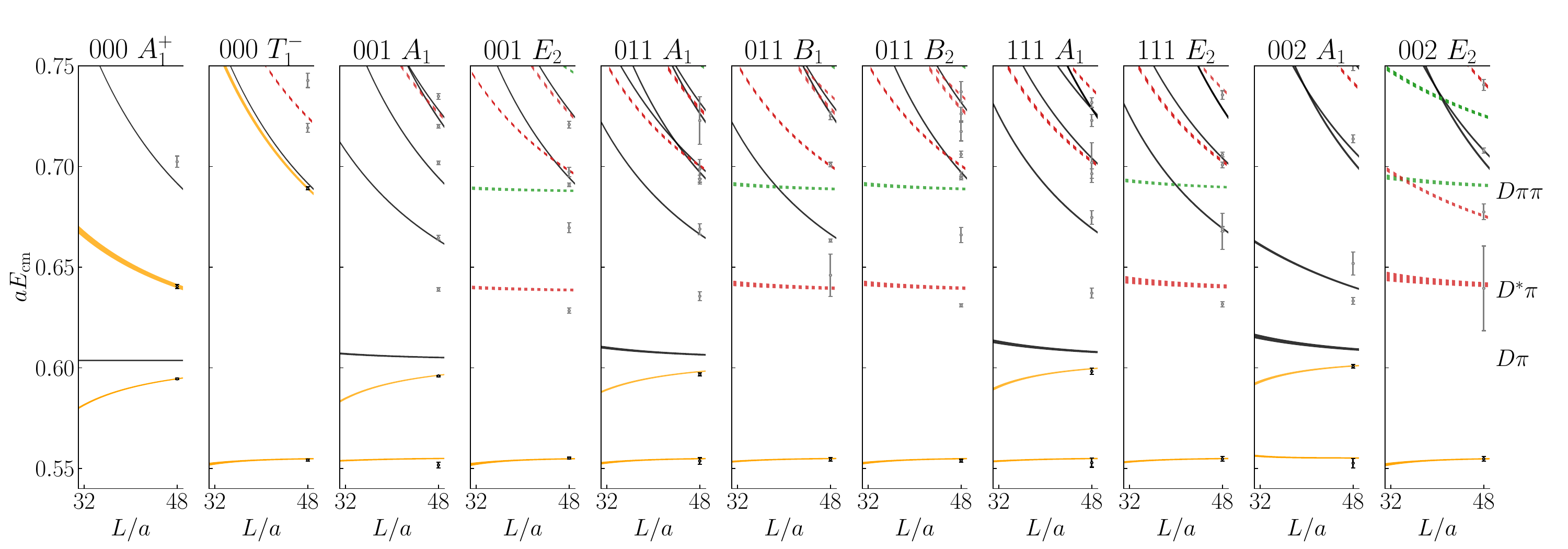}
\caption{The finite-volume spectra at $m_{\pi} \approx 317$ MeV for irreps with leading partial waves being $S$- or $P$-wave. The black and gray points are the energy levels. The black lines, red dashed, and green dashed lines represent the non-interacting energies of the $D\pi$, $D^*\pi$, and the $D\pi\pi$, respectively. Only black points are used in the scattering analysis. The orange bands are the solutions of the L\"{u}scher's equations.}
\label{fig:spectra-Dpi-HH-1d2}
\end{figure*}

\begin{figure*}[htbp]
\includegraphics[width=\textwidth]{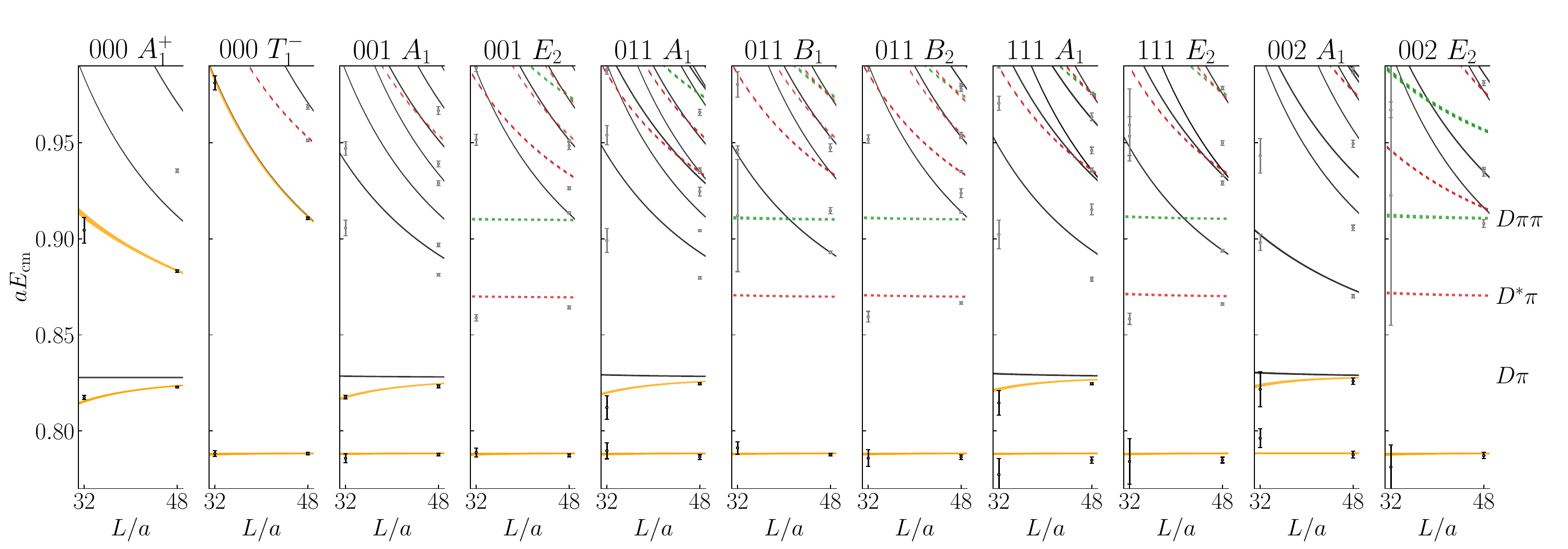}
\caption{$I=\frac{1}{2}$ $D\pi$ finite-volume spectra at $m_{\pi} \approx 208$ MeV. Descriptions as in Fig.~\ref{fig:spectra-Dpi-HH-1d2}.}
\label{fig:spectra-Dpi-FL-1d2}
\end{figure*}

\begin{figure*}[htbp]
\includegraphics[width=\textwidth]{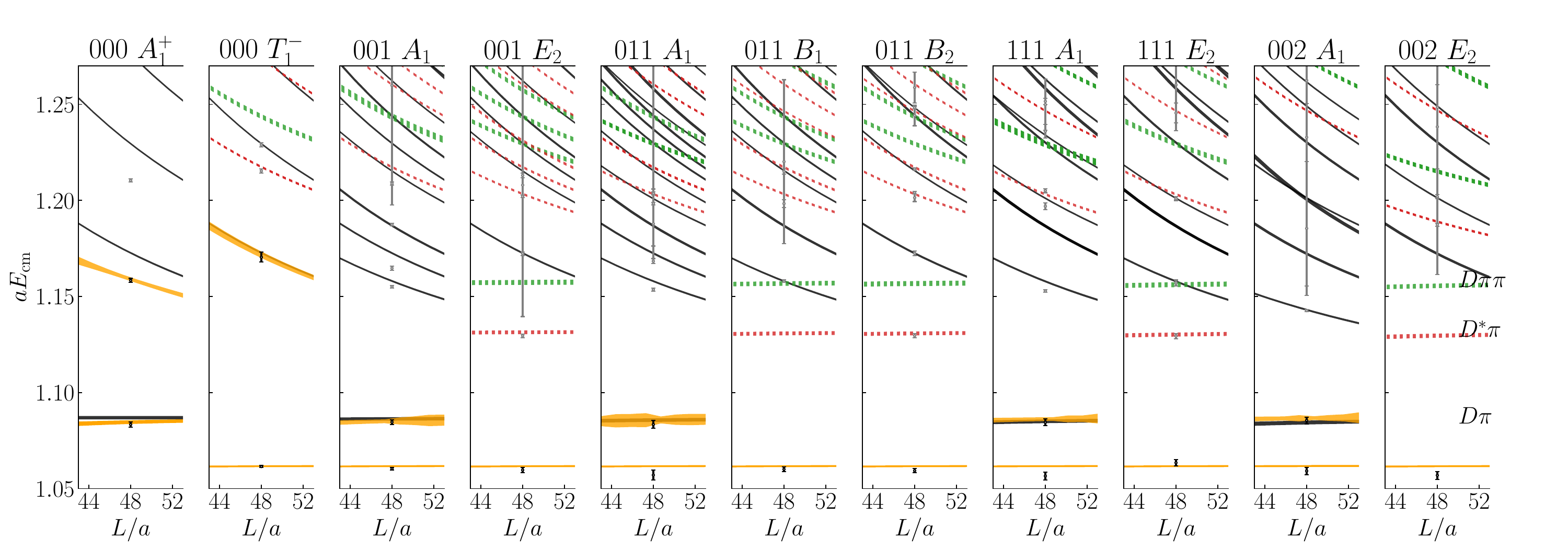}
\caption{$I=\frac{1}{2}$ $D\pi$ finite-volume spectra at $m_{\pi} \approx 133$ MeV. Descriptions as in Fig.~\ref{fig:spectra-Dpi-HH-1d2}.}
\label{fig:spectra-Dpi-CP-1d2}
\end{figure*}

The lattice spectra and the predictions from the L\"{u}shcer's analysis for $I=\frac{1}{2}$ $D\pi$ scattering at $m_{\pi} = 317, 208$, and $133$ MeV are shown in Fig.~\ref{fig:spectra-Dpi-HH-1d2}, \ref{fig:spectra-Dpi-FL-1d2}, \ref{fig:spectra-Dpi-CP-1d2}, respectively. The fact that the first few non-interacting levels of $m_{\pi} = 133$ MeV increase as $L$ increases is a reflection of the discretization artifact since the configuration is coarser than the others.

\subsection{Discretization errors}
\label{sec:supp:discretization}
The dispersion relations are examined and well-described via $E_X = \sqrt{m_X^2+Z_X\vec{k}^2}$, where $\vec{k}$ is the scattering momentum. We find $Z_\pi$ of all configurations is close to $1$. However, $Z_D$ and $Z_{D^*}$ is statistically smaller than $1$, indicating a non-negligible discretization effect. The extracted value of the meson masses, $Z_D$ and $Z_{\pi}$ are tabulated in Tab.~\ref{tab:disp}.

\begin{table}[htbp]
\caption{The meson mass and coefficients in the dispersion relation of the configurations.}
\begin{tabular}{cccccccc}
\toprule
configurations & $a$/fm & $m_{\pi}$/MeV & $m_D$/MeV & $m_{D^*}$/MeV & $Z_{\pi}$ & $Z_D$ & $Z_{D^*}$ \\
\midrule
C48P14 & $0.10530(18)$ & $133.1(1.6)$ & $1903.73(34)$ & $1988.15(72)$ & $1.0103(33)$ & $0.861(14)$ & $0.865(11)$ \\
F32P21 & $0.07746(18)$ & $206.8(2.1)$ & $1901.3(1.1)$ & $2007.2(2.2)$ & $1.0409(84)$ & $0.9283(48)$ & $0.905(16)$ \\
F48P21 & $0.07746(18)$ & $208.12(70)$ & $1900.71(48)$ & $2006.3(1.2)$ & $1.0157(49)$ & $0.9302(31)$ & $0.9360(45)$ \\
F32P30 & $0.07746(18)$ & $305.81(71)$ & $1965.7(1.0)$ & $2076.6(1.5)$ & $0.9977(56)$ & $0.9083(74)$ & $0.9027(93)$ \\
F48P30 & $0.07746(18)$ & $304.98(50)$ & $1966.20(57)$ & $2074.1(1.1)$ & $1.0054(24)$ & $0.9262(41)$ & $0.9243(70)$ \\
H48P32 & $0.05187(26)$ & $317.00(68)$ & $1979.41(85)$ & $2108.9(2.0)$ & $1.0115(41)$ & $0.9551(66)$ & $0.935(15)$ \\
\bottomrule
\end{tabular}
\label{tab:disp}
\end{table}

The coefficients $Z_X$ show more deviation from $1$ at coarser configurations and less at finer configurations, which reflects the degree of the discretization artifact. The discretization error is addressed by comparing the difference between the corrected dispersion relation and the continuum dispersion relation. In other words, the difference of the observables by the $Z_X$ measured on the lattice and $1$ in
\begin{equation}
    E(\vec{k})=\sqrt{m_D^2+Z_D\vec{k}^2} + \sqrt{m_\pi^2+Z_\pi\vec{k}^2}.
\label{eq:disp}
\end{equation}
The errors are found to be small and are added in quadrature with the parametrization errors.

\subsection{Fitting details and parametrization variations}
\label{sec:supp:parametrizations}
The determinant in
\begin{equation}
\operatorname{det}[M(E, \vec{P} ; L)-\cot \delta(E)]=0
\end{equation}
contains an infinite number of partial waves. Since we mainly focus on the energy region near the threshold where the $l$-wave is suppressed by $k^{2l}$, the high partial waves can be safely discarded and the equation then becomes finite-dimensional. The partial wave cutoff is taken to be $P$-wave in this work.

After the cutoff, the L\"{u}scher's equation is still an under-constrained problem. One way to proceed is to parameterize the scattering phase shifts in terms of models and extract the model parameters from the lattice spectra. 

With appropriate parametrization of the scattering phase shifts, the lattice spectra below the inelastic threshold in any irrep can be solved from L\"{u}scher's formula given the specific parameters. The explicit form with an angular momentum cutoff and specific CM momentum can be found in Ref.~\cite{Goeckeler2012}. What we need to solve is the opposite: given the lattice spectra in many irreps, the parameters are determined by minimizing the $\chi^2$, as in Ref.~\cite{Dudek2012},
\begin{equation}
\chi^2(\{\vec{a}\}) = \sum_L \sum_{\substack{\vec{P} \Gamma n \\ \vec{P}^{\prime} \Gamma^{\prime} n^{\prime}}} [ E_{\text{CM}}(L;\vec{P} \Gamma n) - E_{\text{CM}}^{\text{det}}(L;\vec{P} \Gamma n;\{\vec{a}\}) ] \operatorname{Cov}^{-1}(L;\vec{P} \Gamma n; \vec{P}^{\prime} \Gamma^{\prime} n^{\prime}) [ E_{\text{CM}}(L;\vec{P}^{\prime} \Gamma^{\prime} n^{\prime}) - E_{\text{CM}}^{\text{det}}(L;\vec{P}^{\prime} \Gamma n^{\prime};\{\vec{a}\}) ],
\end{equation}
where $\vec{a}$ represents the list of the parameters that need to be determined. $\operatorname{Cov}(L;\vec{P} \Gamma n; \vec{P}^{\prime} \Gamma^{\prime} n^{\prime})$ is the covariance matrix between the energy levels. Note that the covariance between different irrep of the same configuration is not zero, but the covariance between different configurations is zero. $E_{\text{CM}} = \sqrt{E^2 - |\vec{P}|^2}$ is the CM energy fitted from the GEVP eigenvalues, while $E_{\text{CM}}^{\text{det}}$ is the solution of the L\"{u}scher's equation given the parameters. 

When accounting for the discretization errors, Eq.~\ref{eq:disp} is used. This corrected dispersion relation moves the non-interacting levels from the singularities of the conventional zeta function. In order not to miss any energy levels, the energy range is divided into small ranges with a step of $0.0005$ to search for the solutions of the L\"{u}scher's equation.

Besides the ERE used in the main context, we also used the K-matrix parametrization, which provides many flexible forms for each partial wave. For elastic scattering, the $t$-matrix is related to the $K$-matrix by
\begin{equation}
t_{l}^{-1}=\frac{1}{(2 k)^{2l}} K_{l}^{-1} +I_{l},
\end{equation}
where the Chew-Mandelstam phase space $I_{l}$ is taken to be $-i \rho$. The $K$-matrix can be parametrized to be
\begin{equation}
K_l(s)=\frac{g_l^{2}}{m_l^2-s}+\sum_n \gamma_{l}^{(n)} s^n
\end{equation}
or
\begin{equation}
K_{l}(s)=\frac{1}{\sum_{n=0} c_{l}^{(n)} s^n}.
\end{equation}
The number of parameters can be varied to give many different parametrizations. The $K$-matrix with pole parametrizations is used to fit the attractive $I=\frac{1}{2}$ channel. We do not apply the RBW parameterization here since the RBW applies only to resonances that are narrow and there are no relevant thresholds or other nearby resonances. The parametrizations that yield pole positions too deep that touch the left-hand cut are eliminated from this work.

By testing various types of parametrizing the phase shifts, the model dependence is measured. As seen from the right panel of Fig.
1 in the main text, there is little systematic error from the parametrization. The parametrizations we have used in this work for $I=\frac{1}{2}$ $D\pi$ are tabulated in Tab.~\ref{tab:para-Dpi-1d2}. The parametrizations that generate poles in the left-hand region or have large unreasonable $\chi^2/\text{d.o.f}$ are not presented and are used.

\begin{table*}[htbp]
\caption{Parametrizations used for $I=\frac{1}{2}$ $D\pi$ scattering analysis.}
\addtolength{\tabcolsep}{6pt}
\begin{tabular}{c|ccc}
\toprule
configurations & $S$-wave parametrizations & $P$-wave parametrizations & $\chi^2/\text{d.o.f}$ \\
\midrule
\multirow{4}{*}{F32P30 / F48P30} & $k \cot \delta_0 = \frac{1}{a_0} + \frac{1}{2} r_0 k^2$ & $k^3 \cot \delta_1 = \frac{1}{a_1} + \frac{1}{2} r_1 k^2$ & $1.10$ \\
& $k \cot \delta_0 = \frac{1}{a_0} + \frac{1}{2} r_0 k^2 + P_{2,0} k^4$ & $k^3 \cot \delta_1 = \frac{1}{a_1} + \frac{1}{2} r_1 k^2$ & $1.14$ \\
& $k \cot \delta_0 = \frac{1}{a_0} + \frac{1}{2} r_0 k^2 + P_{2,0} k^4$ & $K_1 = \frac{g_1^2}{m_1^2 -s}$ & $1.55$ \\
& $K_0 = \frac{g_0^2}{m_0^2 -s}$ & $k^3 \cot \delta_1 = \frac{1}{a_1} + \frac{1}{2} r_1 k^2$ & $1.10$ \\
\midrule
\multirow{4}{*}{F32P21 / F48P21} & $k \cot \delta_0 = \frac{1}{a_0} + \frac{1}{2} r_0 k^2$ & $k^3 \cot \delta_1 = \frac{1}{a_1} + \frac{1}{2} r_1 k^2$ & $1.94$ \\
& $k \cot \delta_0 = \frac{1}{a_0} + \frac{1}{2} r_0 k^2 + P_{2,0} k^4$ & $k^3 \cot \delta_1 = \frac{1}{a_1} + \frac{1}{2} r_1 k^2$ & $1.76$ \\
& $k \cot \delta_0 = \frac{1}{a_0} + \frac{1}{2} r_0 k^2$ & $K_1 = \frac{g_1^2}{m_1^2 -s}$ & $2.01$ \\
& $k \cot \delta_0 = \frac{1}{a_0} + \frac{1}{2} r_0 k^2 + P_{2,0} k^4$ & $K_1 = \frac{g_1^2}{m_1^2 -s}$ & $1.82$ \\
\midrule
\multirow{2}{*}{H48P32} & $k \cot \delta_0 = \frac{1}{a_0} + \frac{1}{2} r_0 k^2$ & $k^3 \cot \delta_1 = \frac{1}{a_1} + \frac{1}{2} r_1 k^2$ & $1.17$ \\
& $k \cot \delta_0 = \frac{1}{a_0} + \frac{1}{2} r_0 k^2 + P_{2,0} k^4$ & $k^3 \cot \delta_1 = \frac{1}{a_1} + \frac{1}{2} r_1 k^2$ & $1.27$ \\
\midrule
\multirow{1}{*}{C48P14} & $k \cot \delta_0 = \frac{1}{a_0} + \frac{1}{2} r_0 k^2$ & $k^3 \cot \delta_1 = \frac{1}{a_1} + \frac{1}{2} r_1 k^2$ & $1.39$ \\
\bottomrule
\end{tabular}
\addtolength{\tabcolsep}{-6pt}
\label{tab:para-Dpi-1d2}
\end{table*}

As discussed in the main context, a virtual state or a resonance state $D_0^*$ is found for the $S$-wave. While for the $P$-wave, a bound state is found for all pion masses and corresponds to the vector $D^*$. No simple movement of the pole position under $m_{\pi}$ is observed. Note that although $D^*$ should be a resonance in nature. In our physical ensemble, the hyperfine splitting $m_{D^*}-m_{D}$ becomes smaller than $m_{\pi}$ due to the lattice artifacts, making $D^*$ still a bound state. This effect suggests that the results of the ensemble lack precision and are not well-determined. However, due to the larger uncertainty in the ensemble, it imposes fewer constraints on the extrapolations, so the final results are not significantly impacted. Moreover, despite the notable qualitative difference in the $P$-wave, this effect should not raise concerns about the determination of the $S$-wave. Although the $D^*$ is close to the threshold and more susceptible to discretization effects, the mass of the $D_0^*$ is much higher and is less affected by these issues.

We have also attempted to fit the spectra separately for $S$- and $P$-wave. We use $[000] A_1^+$ to fit the $S$-wave and $[000] T_1^-$ to fit the $P$-wave. The results are consistent with that of the combined fit described above.

\subsection{Chiral interpolations of $a_0^{-1}$}
\label{sec:supp:chiral}
\begin{figure}[htbp]
\centering
\includegraphics[width=0.5\columnwidth]{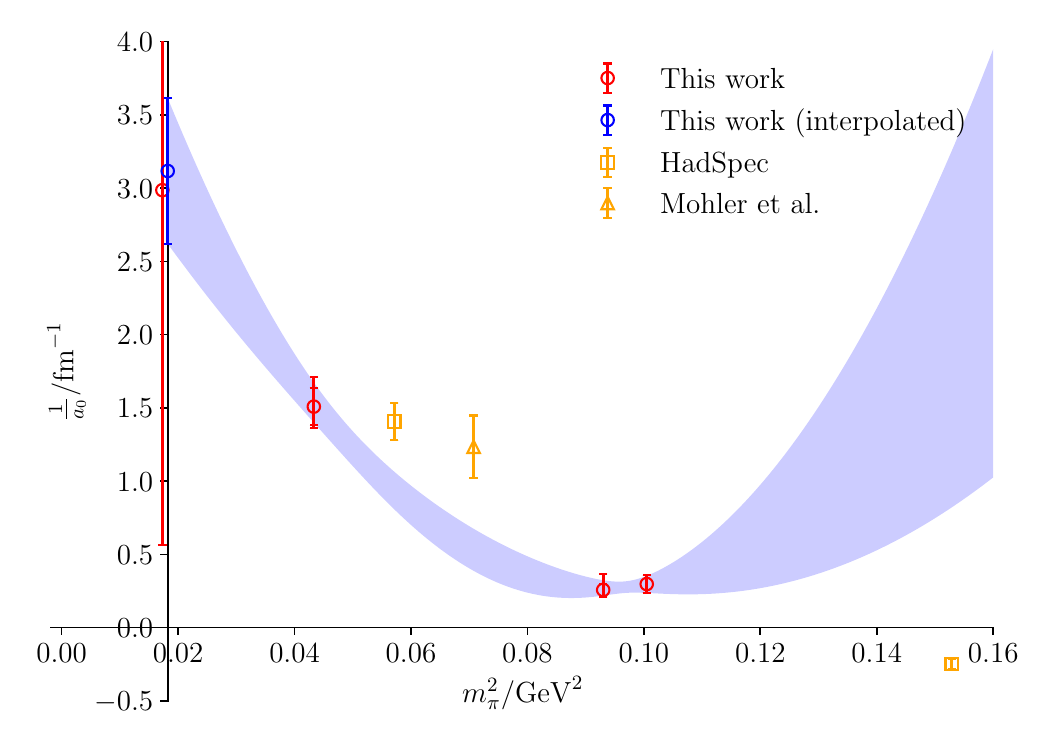}
\caption{The chiral interpolation of the inverse of the $I=\frac{1}{2}$ $D\pi$ scattering length. The fit takes data from this work, shown as the red data. The orange points are from previous lattice works~\cite{Mohler2013, Moir2016, Gayer2021}. The blue band is the reconstructed fitting model. The blue point is the prediction of $\frac{1}{a_0}$ at the physical pion mass.}
\label{fig:interpolation-Dpi-1d2}
\end{figure}

The fit using Eq.
2 in the main text is shown in Fig.~\ref{fig:interpolation-Dpi-1d2}, where the results from other lattice results in Ref.~\cite{Mohler2013, Moir2016, Gayer2021} are also shown. The blue band is the reconstructed curve using the fitted parameters. The blue point is the extrapolated value at the physical pion mass. The extrapolation proceeds by first sampling $a_0^{-1}$ evenly from its distribution for each parametrization at each $m_{\pi}$, followed by performing extrapolations for each sample. This approach incorporates the distribution of all parametrizations across all $m_{\pi}$ values into the fit.

\end{document}